\documentclass[aps,pra,twocolumn,superscriptaddress,floatfix,showpacs]{revtex4-1}

\usepackage{graphicx,graphics}
\usepackage{dcolumn}
\usepackage{amsmath,amssymb,amsfonts}
\usepackage{latexsym,verbatim}
\usepackage{color}
\usepackage{subfigure}
\usepackage[percent]{overpic}
\usepackage{graphicx}
\usepackage{verbatim}

\begin{document}
\title{Exclusive Hong-Ou-Mandel zero-coincidence point}

\author{Yu Yang}
\affiliation{School of Aerospace Science and Technology, Xidian University, Xi'an 710126, China}
\affiliation{Scuola Normale Superiore, I-56126 Pisa, Italy}

\author{Luping Xu}
\email{Corresponding Author: xidian\_lpx@163.com}
\affiliation{School of Aerospace Science and Technology, Xidian University, Xi'an 710126, China}

\author{Vittorio Giovannetti}
\affiliation{NEST, Scuola Normale Superiore and Istituto Nanoscienze-CNR, I-56127 Pisa, Italy}

\begin{abstract}
A generalized multi-parameter Hong-Ou-Mandel interferometer is presented which extends the conventional ``Mandel dip" configuration to the case where a symmetric biphoton source is used to monitor the contemporary presence of $k$ independent time-delays. Our construction results in a two-input/two-output setup, obtained by concatenating 50:50 beam splitters with a collection of adjustable achromatic wave-plates. For $k=1,2$ and $k=4$ explicit examples can be exhibited that prove the possibility of uniquely linking the zero value of  the coincidence counts registered at the output of the interferometer, with the contemporary absence of all the time-delays. Interestingly enough the same result cannot be extended to $k=3$. Besides, the sensitivity of the interferometer is analyzed when the time-delays are affected by the fluctuations over time-scales that are larger than the inverse of the frequency of the pump used to generate the biphoton state. 
\end{abstract}


\maketitle

\section{Introduction}
The Michelson interferometer~\cite{MI1,MI2}, the Hong-Ou-Mandel (HOM) interferometer~\cite{HOM1,HOM2,HOM3,HOM4,HOM5}, and the Mach-Zehnder interferometer (MZI)~\cite{MZI1,MZI2,MZI3,MZ04,MZ01,MZ02,MZ1} are examples of two-input/two-output set-ups which have been extensively used to study two-photon quantum interference effects, with applications in parameter estimation problems, such as phase estimation in the quantum radar~\cite{QR}, or coordinates estimation in quantum positioning system~\cite{Bahder:2004}. In these schemes  a minimum (or a maximum) in the coincidence counts recorded at the output of the device is typically associated with the case where no relative delays affect the propagation of the photons along the two optical paths of the set-up. In the HOM interferometer this correspondence yields the celebrated ``Mandel dip" where, given a symmetric input biphoton (BP) source~\cite{BIPHOTON0,BP-CP-STATE,BIPHOTON1,BIPHOTON2,BIPHOTON3}, a zero-coincidence signal can be uniquely linked to the absence of asymmetries in the signal propagation. Generalization of this effect to more than one parameter are naturally provided by MZIs~\cite{MZI1,MZI2,MZI3,MZ04,MZ01,MZ02,MZ1} where, exploiting the presence of two 50:50 BS, one can in principle monitor two independent time-delays with a single coincidence measurement. It turns out however that for these settings the zero-coincidence event does not exclusively correspond to the contemporary absence of the two delays unless~\cite{MHOM} one includes the presence of an achromatic quarter wave-plate~\cite{ACHRO1,ACHRO2,ACHRO3}. Unlike the standard wave-plates, this optical element provides a constant phase shift independent with the wavelength of the incoming light, typically achieved by using two different birefringent crystalline materials balancing the relative shift in retardation over the wavelength range. As shown in Ref.~\cite{MHOM} by inserting it inside the MZI, one can effectively force an exact swap between the symmetric and anti-symmetric components of the spectral wave function of the propagating biphoton signal, restoring the one-to-one correspondence between the HOM zero-coincidence  point event and the contemporary absence of the delays in the configuration.

Aim of the presented paper is to study the possibility of extending  this result to the case of $k> 2$ time-delay parameters. More specifically, we consider a generalized two-input/two-output interferometer formed by $k$ concatenated 50:50 BSs and $k$ independent time-delay parameters $\tau_1$, $\tau_2$, $\cdots$, $\tau_k$, where with the help of a collection of properly setting achromatic phase-shifts, we try to identify what we dub an exclusive HOM zero-coincidence point event, i.e. a one-to-one correspondence between the zero value in the coincidence counts registered at the output of the interferometer and the contemporary absence of all the time-delays in the scheme. After stating this problem in the rigorous mathematical terms, we observe that while it is explicitly solvable for $k=1,2$ and $4$ (the solutions for $k=1$ and $k=2$ being associated with the results of Refs.~\cite{HOM1} and~\cite{MHOM} respectively), it admits no solution for $k=3$, a peculiar behavior which is probably associated with some accidental symmetries. In the second part of the manuscript we study the sensitivity of the scheme in the presence of random fluctuations with respect to the time-delay parameters $\tau_1$, $\cdots$, $\tau_k$, showing that the effectiveness of the achromatic phase-shifts is strongly affected by such noisy events.

The manuscript is organized as follows: in Sec.~\ref{SEC-GHOM} we introduce the setup, setting the notations and introduce a necessary and sufficient condition for the existence of an exclusive HOM zero-coincidence point for the case of $k$ time-delay parameters. In Sec.~\ref{sec:dip} we hence specify our attentions to $k=3$ and $k=4$ showing that in the first case no solution can be find and presenting instead explicit solution for the second. Moreover some comments for the $k>4$ case is proposed to complete our discussions. In Sec.~\ref{sec:sense} we finally study the sensitivity of the scheme under fluctuating the time-delay parameters. This manuscript finally ends with Sec.~\ref{CONCS} where we present our conclusions and comments on the possible applications to sensing procedures.

\section{Exclusive HOM zero-coincidence point}\label{SEC-GHOM}
Here we propose the scheme and introduce the related notations. More importantly a formal definition of the exclusive HOM zero-coincidence point is presented. 
\subsection{Scheme structure}
\begin{figure*}[!t]
	\centering
	\resizebox{0.8\textwidth}{!}{\includegraphics{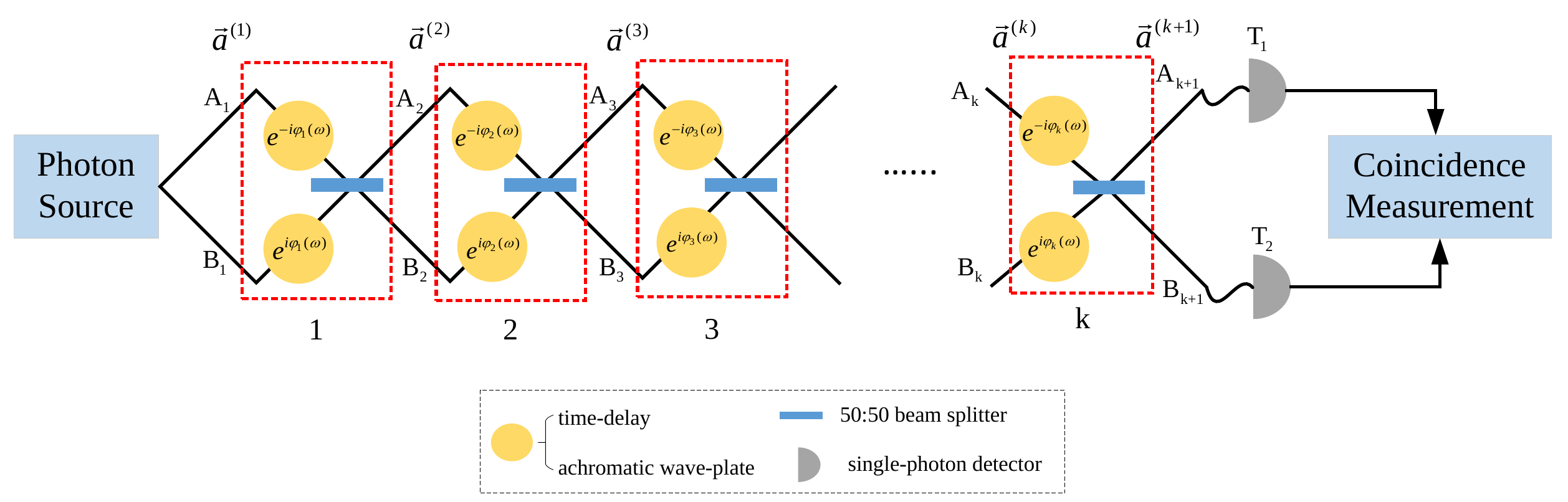}}
	\caption
	{(Color online) The sketch of a generalized interferometer. The yellow circles in the figure represent the phase-shifter elements indicating that light beams propagating on the upper  $\mathbf{A}_\ell$ (lower $\mathbf{B}_\ell$) optical paths, experience the phase-shift $e^{-i\varphi_\ell(\omega)}$ (resp. $e^{i\varphi_\ell}(\omega)$) composed by a time-delay component $\omega \tau_\ell$ ($-\omega\tau_\ell$) and an achromatic wave-plate $\theta_{\ell}/2$ ($-\theta_{\ell}/2$), as indicated by Eq.~(\ref{PHASESHIFT}).} 
	\label{FIGHOM} 
\end{figure*}

Consider the two-input/two-output ports device shown in Fig.~\ref{FIGHOM} which registers the coincidence events at the detectors $T_1$ and $T_2$ associated with a frequency correlated, symmetric biphoton (BP) source~\cite{BIPHOTON0,BP-CP-STATE,BIPHOTON1,BIPHOTON2,BIPHOTON3}. As schematically depicted in Fig.~\ref{FIGHOM}, the setup is obtained by concatenating $k$ modules (red dashed rectangles),  labelled by the progressive index $\ell=1,2, \cdots, k$, each containing the optical elements that introduce the opposite  phase-shifts $e^{-i\varphi_\ell(\omega)}$ and $e^{i\varphi_\ell(\omega)}$  in the lower $\mathbf{A}_\ell$ and upper $\mathbf{B}_\ell$ paths in this generalized interferometer respectively, and a 50:50 beam splitter (BS) that coherently mixes them. Reviewing the original HOM configuration, the phase-shifts  $\varphi_1(\omega)$, $\varphi_2(\omega)$, $\cdots$, $\varphi_k(\omega)$ are assumed to be linked  to the time-delays $\tau_1$, $\tau_2$, $\cdots$, $\tau_k$, which in the following will be treated as the independent variables. Furthermore along the lines detailed in Ref.~\cite{MHOM} we also allow for the presence of achromatic wave-plates~\cite{ACHRO1,ACHRO2,ACHRO3} that add the frequency independent contributions $\theta_{1}$, $\theta_{2}$, $\cdots$, $\theta_{k}\in [0, 2\pi)$ that we shall use as tunable knobs of the device, writing  
\begin{eqnarray} \label{PHASESHIFT} 
\varphi_1(\omega) =   \omega\tau_1 \;, \qquad  \varphi_\ell(\omega) =   \omega\tau_\ell+{\theta_{\ell}}/{2}\;, \quad  \forall \ell \geq 2 \;,
\end{eqnarray} 
the value of $\theta_1$  being  set equal to zero without the loss of generality as it introduces an irrelevant global phase to the final state of the emerging photons, see below. 
Under the above premises, the aim of our analysis is to verify whether it is possible to identify what we dub an exclusive HOM zero-point configuration, i.e. a  special assignment $\bar{\theta}_2, \cdots, \bar{\theta}_k$ of  the parameters $\theta_2, \cdots, \theta_k$  capable to ensuring that a null value for the coincidence counts at $T_1$ and $T_2$ uniquely corresponds to the case where all the temporal delays of the setup are exactly equal to zeros, i.e. 
\begin{eqnarray} \label{RBP0} 
R^{(\bar{\theta}_2, \cdots, \bar{\theta}_k)}_{\mbox{\tiny{BP}}}(\tau_1,...,\tau_k) &=& 0  \nonumber\\
\Longleftrightarrow  \tau_1&=& \tau_2 = \cdots = \tau_k =0\;.
\end{eqnarray} 
For $k=1$ a solution of the above problem is provided by the ``Mandel dip"~\cite{HOM1}. For $k=2$ instead the existence of an exclusive HOM zero-coincidence point follows by the results of  Ref.~\cite{MHOM} which achieves (\ref{RBP0}) by employing  $\bar{\theta}_2= \pi/2$. In what follows, we shall extend this construction to the larger values of $k$: interestingly enough we observe that for $k=3$ no solution can be found, while for $k=4$ special choices of  $\theta_2, \cdots, \theta_4$ exist so that (\ref{RBP0}) holds true.

\subsection{A necessary and sufficient condition for the symmetric BP state}
To set the problem in the rigorous mathematical terms, let us introduce the annihilation operators $\hat{a}_j^{(\ell)}(\omega)$ describing a photon of frequency $\omega$ that enters the $\ell$-th module of the device along the input path $j$ ($j=1(2)$  denotes  the path $\mathbf{A}_\ell$ ($\mathbf{B}_\ell$) respectively) and fulfilling Canonical Commutation Rules (CCR): $[\hat{a}^{(\ell)}_j(\omega),\hat{a}^{(\ell)}_{j'}(\omega^\prime)]=0$, $[\hat{a}^{(\ell)}_j(\omega),\hat{a}^{(\ell)\dagger}_{j^\prime}(\omega^\prime)]=\delta_{j,j^\prime} \delta(\omega-\omega^\prime)$ where $\delta_{j,j^\prime}$ and $\delta(\omega-\omega^\prime)$ being the Kronecker and Dirac deltas respectively. The associated output counterpart $\hat{a}_j^{(\ell+1)}(\omega)$ that is also the input bosonic annihilation operator in the $(\ell+1)$-th module, is connected with $\hat{a}_j^{(\ell)}(\omega)$ by the $\ell$-th module via the following linear transformation 
\begin{eqnarray}
\vec{a}^{(\ell)}(\omega):= \left(\begin{array}{c}
\hat{a}^{(\ell)}_{1}(\omega)\\
\hat{a}^{(\ell)}_{2}(\omega)
\end{array}\right)=M_\ell(\omega) \vec{a}^{(\ell+1)}(\omega)\;,
\end{eqnarray}
with
\begin{eqnarray}
M_\ell(\omega)=\frac{1}{\sqrt{2}}\left( \begin{array}{cc}
e^{i\varphi_\ell(\omega)} & e^{i\varphi_\ell(\omega)}\\
e^{-i\varphi_\ell(\omega)} & -e^{-i\varphi_\ell(\omega)}
\end{array}\right)\;,
\end{eqnarray}
where $2\times 2$ matrix $M_\ell(\omega)$ being defined by the phase-shifts introduced in Eq.~(\ref{PHASESHIFT}). Therefore, the input-output mapping from the first module to the $k$-th module can now be expressed in the compact form 
\begin{eqnarray}\label{MAP}
\vec{a}^{(1)}(\omega)=N_k(\omega) \vec{a}^{(k+1)}(\omega)\;, 
\end{eqnarray}
with the matrix $N_k(\omega)$ defined as 
\begin{eqnarray} \label{DEFNk} 
N_k(\omega)&:=&\left( \begin{array}{cc}
A_k(\omega) & B_k(\omega)\\
C_k(\omega) & D_k(\omega)
\end{array}\right)\nonumber\\
&=&M_1(\omega) M_2(\omega)...M_k(\omega)\;.
\end{eqnarray}

Consider hence  the following  frequency-correlated biphoton pure state \cite{BIPHOTON0,BP-CP-STATE,BIPHOTON1,BIPHOTON2,BIPHOTON3} as the input state of the setup 
\begin{eqnarray}\label{IN}
|\Psi^{(1)}\rangle:=\int d\omega \int d\omega^\prime \Psi(\omega,\omega^\prime)\; \hat{a}_1^{(1)\dagger}(\omega) \hat{a}_2^{(1)\dagger}(\omega^\prime)|\O\rangle\;,
\end{eqnarray}
where $|\O\rangle$ is the multi-mode vacuum state, and where  $\Psi(\omega,\omega^\prime)$ represents a complex amplitude probability distribution on which for the moment we make no assumptions apart from the normalization condition $\int d\omega \int d\omega^\prime |\Psi(\omega,\omega^\prime)|^2 =1$. 
Following the principle of coincidence measurement \cite{MANDELBOOK}, we express the coincidence counts as 
\begin{widetext}
\begin{eqnarray}\label{THEORY}
R^{({\theta}_2, \cdots, {\theta}_k)}_{\mbox{\tiny{BP}}}(\tau_1,...,\tau_k)&:=&\int dt_1 \int dt_2 \langle \Psi^{(k+1)}| \hat{E}_1^{(-)}(t_1)\hat{E}_2^{(-)}(t_2)\hat{E}_2^{(+)}(t_2)\hat{E}_1^{(+)}(t_1)|  \Psi^{(k+1)}\rangle \nonumber\\ 
&=&\int d\omega \int d\omega' \langle \Psi^{(k+1)}| \hat{a}_1^{(k+1)\dagger}(\omega)\hat{a}_1^{(k+1)}(\omega) \hat{a}_2^{(k+1)\dagger}(\omega^\prime)\hat{a}_2^{(k+1)}(\omega^\prime) |\Psi^{(k+1)}\rangle\;,
\end{eqnarray}
\end{widetext}
where $\hat{E}_j^{(-)}(t)=(\hat{E}_j^{(+)})^\dagger:=\frac{1}{\sqrt{2\pi}}\int d\omega \hat{a}_j^{(k+1) \dagger}(\omega)e^{i\omega t}$  is the amplitude of electromagnetic field at detector $T_j$,  and where $|\Psi^{(k+1)}\rangle$ represents the output state  emerging from the interferometer associated with the input state  $|\Psi^{(1)}\rangle$. The former one can be obtained by using  (\ref{MAP}) to express $\hat{a}_j^{(1)\dagger}(\omega)$ in terms of the corresponding output-mode operator $\hat{a}_j^{(k+1) \dagger}(\omega)$ obtaining  
\begin{eqnarray}\label{MATRIXMAP}
\hat{a}_1^{(1)\dagger}(\omega)=A_k^*(\omega)\hat{a}_1^{(k+1)\dagger}(\omega)+B^*_k(\omega)\hat{a}_2^{(k+1)\dagger}(\omega)\;, \nonumber\\ 
\hat{a}_2^{(1)\dagger}(\omega)=C_k^*(\omega)\hat{a}_1^{(k+1)\dagger}(\omega)+D^*_k(\omega)\hat{a}_2^{(k+1)\dagger}(\omega)\;,
\end{eqnarray}
which is inserted  into (\ref{IN}) gives us
\begin{eqnarray}\label{OUTO}
&|\Psi^{(k+1)}\rangle&=\int d\omega \int d\omega^\prime \Psi(\omega,\omega^\prime)\nonumber\\ &&\times[A_k^*(\omega)\hat{a}_1^{(k+1)\dagger}(\omega)+B_k^*(\omega)\hat{a}^{(k+1)\dagger}_2(\omega)]\nonumber\\
&&\times[C_k^*(\omega^\prime)\hat{a}_1^{(k+1)\dagger}(\omega^\prime)+D_k^*(\omega^\prime)\hat{a}^{(k+1)\dagger}_2(\omega^\prime)]|\O\rangle\;.\nonumber\\
\end{eqnarray}
Expanding Eq.(\ref{OUTO}) we observe that it contains two kinds of contributions: 
the first contains the terms where both photons belong to a same output port of the interferometer (either $\mathbf{A}_{k+1}$ or $\mathbf{B}_{k+1}$) and gives explicitly no contribution to (\ref{THEORY});
the second instead contains all the terms where one photon in $\mathbf{A}_{k+1}$ and another one in $\mathbf{B}_{k+1}$ and which can actively contribute to 
$R^{({\theta}_2, \cdots, {\theta}_k)}_{\mbox{\tiny{BP}}}(\tau_1,...,\tau_k)$. Its analytic expression is given by 
\begin{eqnarray}\label{OUT}
|\Phi^{(k+1)}\rangle&=&\int d\omega \int d\omega^\prime \Phi^{(k+1)}(\omega,\omega^\prime) \nonumber\\
&&\times \hat{a}_1^{(k+1)\dagger}(\omega) \hat{a}^{(k+1)\dagger}_2(\omega^\prime) |\O\rangle \;, 
\end{eqnarray}
where $\Phi^{(k+1)}(\omega,\omega^\prime)$ is the new biphoton amplitude that we can write as
\begin{eqnarray} 
&\Phi^{(k+1)}&(\omega,\omega^\prime)  \nonumber\\
&:=&\Psi(\omega,\omega^\prime) A_k^*(\omega)D_k^*(\omega^\prime) + \Psi(\omega^\prime,\omega)B_k^*(\omega^\prime)C_k^*(\omega)\;, \nonumber \\
&:=&\Psi_S(\omega,\omega^\prime) \mbox{Perm}_k^{*}(\omega,\omega') +\Psi_A(\omega,\omega^\prime) \mbox{Det}_k^{*}( \omega,\omega')\;,\nonumber\\ 
\end{eqnarray} 
where in the last line we use the symmetric and antisymmetric components of the input distribution
\begin{eqnarray} 
\Psi_S(\omega,\omega^\prime):=( \Psi(\omega,\omega^\prime)+\Psi(\omega^\prime,\omega))/2\;, \nonumber\\
\Psi_A(\omega,\omega^\prime):=( \Psi(\omega,\omega^\prime)-\Psi(\omega^\prime,\omega))/2\;,
\end{eqnarray} 
and introduce the functions 
\begin{eqnarray}  \label{PERM} 
\mbox{Perm}_k(\omega,\omega') &:= & A_k(\omega)D_k(\omega^\prime) + B_k(\omega^\prime)C_k(\omega) \;, \nonumber\\
\mbox{Det}_k( \omega,\omega') &:= & A_k(\omega)D_k(\omega^\prime) - B_k(\omega^\prime)C_k(\omega)\;,
\end{eqnarray} 
that correspond respectively to the permanent~\cite{BATHIA}  and determinant of the $2\times 2$ matrix
$N_k(\omega,\omega') :=\left( \begin{array}{cc}
A_k(\omega) & B_k(\omega')\\
C_k(\omega) & D_k(\omega')
\end{array}\right)$ and which exhibit an implicit dependence upon the delays $\tau_1, \cdots, \tau_k$ and upon the constant phase shifts $\theta_2$, $\cdots$, $\theta_k$. Replacing all this into Eq.~(\ref{THEORY}) we finally get
\begin{widetext}
\begin{eqnarray}\label{BP}
&&R^{({\theta}_2, \cdots, {\theta}_k)}_{\mbox{\tiny{BP}}}(\tau_1,...,\tau_k)=\langle\Phi^{(k+1)}|\Phi^{(k+1)}\rangle = \int d\omega \int  d\omega^\prime \; \Big|\Phi^{(k+1)} (\omega,\omega^\prime)\Big|^2\nonumber\\
&&\qquad =\frac{1}{4} \int d\omega \int  d\omega^\prime \; \Big| \Psi_S(\omega,\omega^\prime) [ \mbox{Perm}^*_k(\omega,\omega') +\mbox{Perm}^*_k(\omega',\omega)]+ \Psi_A(\omega,\omega^\prime) [ \mbox{Det}^*_k(\omega,\omega') - \mbox{Det}^*_k(\omega',\omega)]\Big|^2  \nonumber \\
&&\;  \qquad + \frac{1}{4} \int d\omega \int  d\omega^\prime \; \Big|\Psi_S(\omega,\omega^\prime) [ \mbox{Perm}^*_k(\omega,\omega') -
\mbox{Perm}^*_k(\omega',\omega)]+ \Psi_A(\omega,\omega^\prime) [ \mbox{Det}^*_k(\omega,\omega') + \mbox{Det}^*_k(\omega',\omega)] \Big|^2 \;,
\end{eqnarray}
\end{widetext} 
where in the second line we separate the symmetric and antisymmetric contributions of $\Phi^{(k+1)} (\omega,\omega^\prime)$. The above expression makes it evident that a zero value of coincidence counts can be obtained if and only if the following conditions get satisfied for all $\omega$ and $\omega'$,
\begin{widetext}
\begin{eqnarray} \label{CONDI1} 
\left\{ \begin{array}{c} 
\Psi_S(\omega,\omega^\prime) [ \mbox{Perm}^*_k(\omega,\omega') +\mbox{Perm}^*_k(\omega',\omega)]+ \Psi_A(\omega,\omega^\prime) [ \mbox{Det}^*_k(\omega,\omega') - \mbox{Det}^*_k(\omega',\omega)] =0 \;, \\  \\ 
\Psi_S(\omega,\omega^\prime) [ \mbox{Perm}^*_k(\omega,\omega') - \mbox{Perm}^*_k(\omega',\omega)]+ \Psi_A(\omega,\omega^\prime) [ \mbox{Det}^*_k(\omega,\omega') + \mbox{Det}^*_k(\omega',\omega)]=0  \;.
\end{array} \right. 
\end{eqnarray} 
In particular under the simplifying hypothesis of an input BP state that has a symmetric amplitude analogous to those analyzed in Ref.~\cite{MHOM}, i.e. 
\begin{eqnarray} 
\Psi_A(\omega,\omega') =0   \Longrightarrow   \Psi(\omega,\omega') = \Psi_S(\omega,\omega') \;, 
\end{eqnarray} 
Eq.(\ref{CONDI1}) implies a simple necessary and sufficient condition for having a zero-coincidence counts, i.e. 
\begin{eqnarray} \label{RBP1} 
R^{({\theta}_2, \cdots,{\theta}_k)}_{\mbox{\tiny{BP}}}(\tau_1,...,\tau_k) = \int d\omega \int  d\omega^\prime  |  \Psi_S(\omega,\omega^\prime) |^2 |\mbox{Perm}_k(\omega,\omega')|^2 =0 
\quad \Longleftrightarrow  \quad
\mbox{Perm}_k(\omega,\omega')=0 \quad  \forall \omega,\omega'\in {\cal D}\;,
\end{eqnarray}
\end{widetext} 
which in the following we shall adopt  to study the problem (\ref{RBP0}) --  ${\cal D}$ being the domain where $\Psi_S(\omega,\omega')$ is supported.

\section{Multi-parameter HOM zero-coincidence point}\label{sec:dip}
From the discussion of  Sec~\ref{SEC-GHOM}, the presence of a zero value  in the coincidence counts $R^{({\theta}_2, \cdots,{\theta}_k)}_{\mbox{\tiny{BP}}}(\tau_1,...,\tau_k)$ when feeding the apparatus with a symmetric biphoton state $|\Psi^{(1)}\rangle$ is related with the possibility of nullifying the function $\mbox{Perm}_k(\omega,\omega')$ for all points in the support ${\cal D}$ of $\Psi_S(\omega,\omega')$ which, without the loss of generality hereafter we shall assume to be the full frequency domain. 
As for the scheme defined by a single modulus ($k=1$), Eqs.~(\ref{DEFNk}) and (\ref{PERM}) reduce to 
\begin{eqnarray} 
N_1(\omega)=M_1(\omega)&=&\frac{1}{\sqrt{2}}\left( \begin{array}{cc}
e^{i\omega\tau_1}&e^{i\omega\tau_1}\\
e^{-i\omega\tau_1}&-e^{-i\omega\tau_1}
\end{array}\right) \;,  \\
\mbox{Perm}_1(\omega,\omega')&=&-i\sin((\omega-\omega^\prime)\tau_1)\;.
\end{eqnarray}
Therefore, $\mbox{Perm}_1(\omega,\omega')=0$ for all $\omega,\omega'\in {\cal D}$ if and only if $\tau_1=0$. Under this assumption we get 
\begin{eqnarray} \label{RBP1e} 
R_{\mbox{\tiny{BP}}}(\tau_1)=\int d\omega \int  d\omega^\prime  |  \Psi_S(\omega,\omega^\prime) |^2 \sin^2((\omega-\omega^\prime)\tau_1)\;, \nonumber\\
\end{eqnarray} 
which corresponds to the standard result of coincidence counts observed in the conventional HOM interferometer~\cite{HOM1} exhibiting $\tau_1=0$ as an exclusive HOM zero-coincidence point (``Mandel dip").

A less non trivial configuration is already obtained in the case of $k=2$ modules which was studied in Ref.~\cite{MHOM}. Here Eqs.~(\ref{DEFNk}) and (\ref{PERM}) yield 
\begin{eqnarray}
&&N_2(\omega)=M_1(\omega) M_2(\omega) \nonumber\\
&&=\left(\begin{array}{cc}
e^{i \omega\tau_1} \cos(\omega \tau_2 + \theta_2/2)  &  i e^{i \omega\tau_1} \sin(\omega \tau_2 + \theta_2/2) \\ \\
i e^{-i \omega\tau_1} \sin(\omega \tau_2 + \theta_2/2) &  e^{-i \omega\tau_1} \cos(\omega \tau_2 + \theta_2/2)
\end{array} \right) \;,  \nonumber\\
\end{eqnarray}
and  
\begin{eqnarray} \label{PERM2} 
\mbox{Perm}_2(\omega,\omega')&=&\cos((\omega-\omega^\prime)\tau_1) \cos((\omega+\omega^\prime)\tau_2+\theta_2)  \nonumber\\
&+&i\sin((\omega-\omega^\prime)\tau_1)\cos((\omega-\omega^\prime)\tau_2)\;,
\end{eqnarray}
which for $\tau_1=\tau_2=0$  gives $\mbox{Perm}_2(\omega,\omega')\Big|_{\tau_1=\tau_2=0}= \cos\theta_2$.
Accordingly from Eq.~(\ref{RBP1}) it follows that we can have $R^{({\theta}_2)}_{\mbox{\tiny{BP}}}(0,0) = 0$
by setting $\theta_2=\bar{\theta}_2=\pi/2$. Most importantly under this condition (\ref{PERM2}) becomes 
\begin{eqnarray}\label{ONEPHASE}
&\mbox{Perm}_2(\omega,\omega')&\Big|_{\theta_2=\pi/2}
=-\cos((\omega-\omega^\prime)\tau_1)  \sin((\omega+\omega^\prime)\tau_2) \nonumber\\
&&+i\sin((\omega-\omega^\prime)\tau_1)\cos((\omega-\omega^\prime)\tau_2)\;,
\end{eqnarray}
for which only $\tau_1=\tau_2=0$ can ensure the fulfillment of Eq.~(\ref{RBP1}). Hence also in this $k=2$ case we can conclude that the scheme exhibits an exclusive zero-coincidence point~(\ref{RBP0}) under the special setting $\theta_2=\pi/2$~\cite{MHOM}. 

\subsection{Absence of the exclusive zero-coincidence point for $k=3$ modules} 
Now we consider the case with respect to $k=3$ modules, under this condition Eqs.(\ref{DEFNk}) and (\ref{PERM}) yield 
\begin{widetext}
\begin{eqnarray}
N_3(\omega)&=&M_1(\omega) M_2(\omega) M_3(\omega) \nonumber\\
&=&\frac{1}{\sqrt{2}} \left( \begin{array}{ccc}
e^{i \varphi_1} [\cos(\varphi_2- \varphi_3) + i  \sin(\varphi_2 + \varphi_3)]& & e^{i \varphi_1} [\cos(\varphi_2+ \varphi_3) - i  \sin(\varphi_2 - \varphi_3)] \\ \\
e^{-i \varphi_1} [\cos(\varphi_2+ \varphi_3) + i  \sin(\varphi_2 - \varphi_3)] & & e^{-i \varphi_1} [-\cos(\varphi_2- \varphi_3) + i  \sin(\varphi_2 + \varphi_3)] \end{array}\right) \;,
\end{eqnarray}
and  
\begin{eqnarray}  \label{PERM3} 
\mbox{Perm}_3(\omega,\omega') &=& -\cos(\varphi_1-\varphi_1')  \sin(\varphi_2+  \varphi'_2) \sin(\varphi_3+\varphi'_3) +  \sin(\varphi_1-\varphi_1')  \sin(\varphi_2-  \varphi'_2) \cos(\varphi_3+\varphi'_3)     \nonumber \\
&& - i\Big( \cos(\varphi_1-\varphi_1')\cos(\varphi_2+  \varphi'_2)\sin(\varphi_3 - \varphi'_3) + \sin(\varphi_1-\varphi_1')\cos(\varphi_2-  \varphi'_2) \cos(\varphi_3-\varphi'_3)\Big)\;,  
\end{eqnarray} 
\end{widetext}
where $\varphi_\ell:=\varphi_\ell(\omega)$, $\varphi_\ell'=\varphi_\ell(\omega')$ for $\ell=1,2,3$.
Recalling~(\ref{PHASESHIFT}) one can easily verify that for  $\tau_1=\tau_2=\tau_3=0$ Eq.~(\ref{PERM3}) reduces to 
\begin{eqnarray} \label{COND333} 
\mbox{Perm}_3(\omega,\omega') \Big|_{\tau_1=\tau_2=\tau_3=0} = -\sin(\theta_2) \sin(\theta_3)\;, 
\end{eqnarray} 
which can be forced to zero by taking one (or both) of the two phase shifts $\theta_2$ and $\theta_3$ equal to an integer multiple of $\pi$.  
Interestingly enough none of these settings provide an  exhaustive zero-coincidence point~(\ref{RBP0}) for the scheme. For instance assuming  $\theta_2=\pi$ we get 
\begin{eqnarray} 
&&\mbox{Perm}_3(\omega,\omega')\Big|_{\theta_2=\pi}  \nonumber \\
&=&\cos((\omega-\omega')\tau_1)  \sin((\omega+\omega')\tau_2)  \sin((\omega+\omega')\tau_3 + \theta_3)  \nonumber \\
&+&\sin((\omega-\omega')\tau_1)  \sin((\omega-\omega')\tau_2)\cos((\omega+\omega')\tau_3 + \theta_3)  \nonumber \\
&+&i\Big( \cos((\omega-\omega')\tau_1)  \cos((\omega+\omega')\tau_2)\sin((\omega-\omega')\tau_3) \nonumber \\
&-&\sin((\omega-\omega')\tau_1)  \cos((\omega-\omega')\tau_2) \cos((\omega-\omega')\tau_3)\Big)\;,\nonumber
\end{eqnarray}
which besides $\tau_1=\tau_2=\tau_3=0$ admits zero value as all the points $(\tau_1,\tau_2,\tau_3)$ proportional to $(1, 0, 1)$ -- in the case where 
$\theta_2=0$ the same hold for $(\tau_1,\tau_2,\tau_3)$ proportional to $(1, 0, -1)$. On the contrary for $\theta_3=\pi$ we have  
\begin{eqnarray} 
&&\mbox{Perm}_3(\omega,\omega')\Big|_{\theta_3=\pi} \nonumber\\
&=&\cos((\omega-\omega')\tau_1)  \sin((\omega+\omega')\tau_2 + \theta_2)  \sin((\omega+\omega')\tau_3)\nonumber\\
&-&\sin((\omega-\omega')\tau_1)  \sin((\omega-\omega')\tau_2)\cos((\omega+\omega')\tau_3) \nonumber\\
&-&i\Big( \cos((\omega-\omega')\tau_1)    \cos((\omega+\omega')\tau_2 + \theta_2)\sin((\omega-\omega')\tau_3)\nonumber\\
&+&\sin((\omega-\omega')\tau_1)  \cos((\omega-\omega')\tau_2) \cos((\omega-\omega')\tau_3)\Big) \;,\nonumber
\end{eqnarray} 
which instead admits zero value for all points $(\tau_1,\tau_2,\tau_3)$ proportional to $(0, 1, 0)$ -- the same result also when  $\theta_3=0$. 

\subsection{Exclusive zero-coincidence point for $k=4$ modules} 
The presence of exclusive zero-coincidence solutions is restored for $k=4$. Analytically Eqs.(\ref{DEFNk}) and (\ref{PERM}) get replaced by
\begin{widetext}
\begin{eqnarray}
&&N_4(\omega)=M_1(\omega)M_2(\omega)M_3(\omega)M_4(\omega) \nonumber\\
&&=\left( \begin{array}{ccc}
e^{i\varphi_1}[\cos(\varphi_3)\cos(\varphi_2+\varphi_4)+i \sin(\varphi_3)\cos(\varphi_2-\varphi_4)] & & e^{i\varphi_1}[\sin(\varphi_3)\sin(\varphi_2-\varphi_4)+i \cos(\varphi_3)\sin(\varphi_2+\varphi_4)]\\ \\
-e^{-i\varphi_1}[\sin(\varphi_3)\sin(\varphi_2-\varphi_4)-i\cos(\varphi_3)\sin(\varphi_2+\varphi_4)] & &
e^{-i\varphi_1}[\cos(\varphi_3)\cos(\varphi_2+\varphi_4)-i\sin(\varphi_3)\cos(\varphi_2-\varphi_4)]
\end{array}\right)\;, \nonumber\\
\end{eqnarray}
and
\begin{eqnarray}\label{4PERM}
\mbox{Perm}_4(\omega,\omega^\prime) &=&\cos(\varphi_1-\varphi_1^\prime)[\cos(\varphi_2+\varphi_2^\prime)\cos(\varphi_3-\varphi_3^\prime)\cos(\varphi_4+\varphi_4^\prime)-\sin(\varphi_2+\varphi_2^\prime)\cos(\varphi_3+\varphi_3^\prime)\sin(\varphi_4+\varphi_4^\prime)]\nonumber\\
&-&\sin(\varphi_1-\varphi_1^\prime)[\cos(\varphi_2-\varphi_2^\prime)\sin(\varphi_3-\varphi_3^\prime)\cos(\varphi_4+\varphi_4^\prime)+\sin(\varphi_2-\varphi_2^\prime)\sin(\varphi_3+\varphi_3^\prime)\sin(\varphi_4+\varphi_4^\prime)]\nonumber\\
&+&i\cos(\varphi_1-\varphi_1^\prime)[\cos(\varphi_2+\varphi_2^\prime)\sin(\varphi_3-\varphi_3^\prime)\cos(\varphi_4-\varphi_4^\prime)+\sin(\varphi_2+\varphi_2^\prime)\sin(\varphi_3+\varphi_3^\prime)\sin(\varphi_4-\varphi_4^\prime)]\nonumber\\
&+&i\sin(\varphi_1-\varphi_1^\prime)[\cos(\varphi_2-\varphi_2^\prime)\cos(\varphi_3-\varphi_3^\prime)\cos(\varphi_4-\varphi_4^\prime)-\sin(\varphi_2-\varphi_2^\prime)\cos(\varphi_3+\varphi_3^\prime)\sin(\varphi_4-\varphi_4^\prime)]\;,\nonumber\\
\end{eqnarray}
\end{widetext}
where adopting the same simplifying notation introduced in Eq.~(\ref{PERM3}). For $\tau_1=\tau_2=\tau_3=\tau_4=0$, the above expression  reduces to
\begin{eqnarray}\label{TOTAL0}
&&\mbox{Perm}_4(\omega,\omega^\prime)\Big|_{\tau_1=\tau_2=\tau_3=\tau_4=0}  \nonumber\\
&&=\cos(\theta_2)\cos(\theta_4)-\sin(\theta_2)\sin(\theta_4) \cos(\theta_3)\;.
\end{eqnarray} 
We notice that taking for instance $(\theta_2,\theta_4) =(0,\pi/2)$ the above expression can be forced to zero. Yet, one can easily verify that under this condition we do not get an exclusive HOM zero-coincidence point as $\mbox{Perm}_4(\omega,\omega^\prime)$ also nullifies for $(\tau_1,\tau_2,\tau_3,\tau_4)$ e.g. proportional to the vector $(1,0,-1,0)$. Similar discussion holds also for the cases where either $\theta_2=0,\pi$ and $\theta_4=\pi/2,3\pi/2$, or where instead $\theta_2=\pi/2,3\pi/2$, $\theta_4=0,\pi$. Excluding these cases, e.g. requiring $\sin\theta_2 \sin\theta_4 \ne 0$, we can still force Eq.~(\ref{TOTAL0}) to zero by fixing the achromatic phases to fulfill the identity 
\begin{eqnarray} \label{QUESTA} 
\theta_3 = \arccos(\cot \theta_2 \cot \theta_4)\;.
\end{eqnarray} 
We claim that under these conditions the model exhibits indeed an exclusive zero-coincidence point. For this purpose we observe that from Eq.~(\ref{RBP1}) it follows that a necessary condition to have zero value for coincidence counts is to nullify the real part of $\mbox{Perm}_4(\omega,\omega')$ for all possible choices of $\omega$ and $\omega'$ in the domain,  i.e. 
\begin{eqnarray} 
R^{({\theta}_2, \cdots,{\theta}_4)}_{\mbox{\tiny{BP}}}(\tau_1,...,\tau_4)&=&0 \nonumber\\
\Longrightarrow   
\mbox{Re}[\mbox{Perm}_4(\omega,\omega')]&=&0 \quad  \forall \omega, \omega' \in {\cal D}\;.
\end{eqnarray} 
Restricting the analysis to the case $\omega=\omega'$ the above equation yields hence the following condition 
\begin{eqnarray}\label{4PERM1new}
&&\mbox{Re}[\mbox{Perm}_4(\omega,\omega) ] =\cos(2 \omega \tau_2+\theta_2)\cos(2 \omega \tau_4+\theta_4)  \nonumber\\
&&-\sin(2 \omega \tau_2+\theta_2)\sin(2 \omega \tau_4+\theta_4) \cos(2 \omega \tau_3+\theta_3) =0 \;,    \nonumber\\
\end{eqnarray}
which must hold for all  $\omega \in {\cal D}$ if we want to get $R^{({\theta}_2, \cdots,{\theta}_4)}_{\mbox{\tiny{BP}}}(\tau_1,...,\tau_4) =0$: as shown in Sec.~\ref{subsect}  below however this condition can only be achieved when  $\tau_2=\tau_3=\tau_4=0$, which when inserted into (\ref{4PERM}) finally leads to identify $\tau_1=\tau_2=\tau_3=\tau_4=0$ as the unique zero-coincidence point of the model. 

\subsubsection{Existence of the exclusive zero-coincidence point} \label{subsect} 
Here we explicitly show that for proper choices of $\theta_2,\theta_3,\theta_4$ as in Eq.~(\ref{QUESTA}), Eq.~(\ref{4PERM1new}) admits $\tau_1=\tau_2=\tau_3=\tau_4=0$ as unique solution, hence proving that for  $k=4$ we can have the exclusive zero-coincidence point condition~(\ref{RBP0}). Firstly, let us observe that we can exclude the cases with $\tau_2=\tau_4=0$. Indeed under the condition of $\tau_2=\tau_4=0$ Eq.~(\ref{4PERM1new}) yields
\begin{eqnarray}\label{4PERM12}
&&\mbox{Re}[\mbox{Perm}_4(\omega,\omega)]=\cos(\theta_2)\cos(\theta_4)  \nonumber\\
&&-\sin(\theta_2)\sin(\theta_4) \cos(2 \omega \tau_3+\theta_3) =0 \;, 
\end{eqnarray}
which, considering that we are discussing the case where $\sin\theta_2\sin \theta_4 \ne 0$ and Eq.~(\ref{QUESTA}) holds true, can be satisfied for all $\omega$ only taking $\tau_3=0$. However now we shall have $\tau_2=\tau_3=\tau_4=0$ so that Eq.~(\ref{4PERM}) becomes
\begin{eqnarray}\label{4PERM22}
\mbox{Perm}_4(\omega,\omega^\prime)=i \sin((\omega-\omega')\tau_1)\;, 
\end{eqnarray}
which is null for all $\omega$ and $\omega'$ only when we have also $\tau_1=0$. Secondly, considering the case where at least one among $\tau_2$ and $\tau_4$ is different from zero. Under this condition it is convenient to rewrite Eq.(\ref{4PERM1new}) as 
\begin{eqnarray} \label{DDF|} 
g_1(\omega) = g_2 (\omega)\;,
\end{eqnarray} 
with 
\begin{eqnarray}\label{4PERM1345}
&g_1(\omega)&: = \cos(2 \omega \tau_2+\theta_2)\cos(2 \omega \tau_4+\theta_4) \nonumber\\
&&= \frac{1}{2} \cos( 2 \omega(\tau_2 + \tau_4) + \theta_2 +\theta_4)  \nonumber\\
&&+ \frac{1}{2} \cos( 2 \omega(\tau_2 - \tau_4) + \theta_2 -\theta_4)\;,
\end{eqnarray}
and 
\begin{eqnarray}\label{4PERM1sdfgsf}
&g_2(\omega)&: = \sin(2 \omega \tau_2+\theta_2)\sin(2 \omega \tau_4+\theta_4) \cos(2 \omega \tau_3+\theta_3)  \nonumber \\
&&=\frac{1}{4} \cos( 2 \omega(\tau_2 - \tau_4+\tau_3) + \theta_2 -\theta_4+\theta_3)  \nonumber \\
&&+\frac{1}{4} \cos( 2 \omega(\tau_2 - \tau_4-\tau_3) + \theta_2 -\theta_4-\theta_3) \nonumber \\ 
&&-\frac{1}{4} \cos( 2 \omega(\tau_2 + \tau_4+\tau_3) + \theta_2 + \theta_4+\theta_3) \nonumber \\
&&-\frac{1}{4} \cos( 2 \omega(\tau_2 + \tau_4-\tau_3) + \theta_2 + \theta_4-\theta_3) \;. 
\end{eqnarray}
We now observe that if at least one among $\tau_2$ and $\tau_4$ is different from zero, then $g_1(\omega)$ is an oscillating function of $\omega$ characterized by two independent frequencies $2\omega(\tau_2 + \tau_4)$ and $2\omega(\tau_2 - \tau_4)$.
On the contrary in general $g_2(\omega)$ admits four different frequencies (explicitly given by $ 2\omega(\tau_2 \pm \tau_4\pm \tau_3)$) apart from the special degenerate cases where we have either $\tau_2 + \tau_4 -  \tau_3=\tau_2 - \tau_4 +  \tau_3$, or $\tau_2 + \tau_4 +  \tau_3=\tau_2 - \tau_4 -  \tau_3$, which both admits 3 independent frequency values. 
The only possibility to fulfill (\ref{DDF|})  (and hence (\ref{4PERM1new})) is that in the above configurations, the multiplicative coefficients of such frequency terms help us to reduce the effective number of frequencies of $g_2(\omega)$ to two. By looking at Eq.~(\ref{4PERM1sdfgsf}) however, this can only happen in the degenerate cases. For instance, for $\tau_2 + \tau_4 -  \tau_3=\tau_2 - \tau_4 +  \tau_3$ by taking $ \theta_2 +\theta_4-\theta_3=\theta_2 -\theta_4+\theta_3$ we can reduce Eq.(\ref{4PERM1sdfgsf}) to 
\begin{eqnarray}\label{4PERM1sdfgsfd}
g_2(\omega)&: =&\frac{1}{4} \cos( 2 \omega(\tau_2 - \tau_4-\tau_3) + \theta_2 -\theta_4-\theta_3)  \nonumber\\
&-& \frac{1}{4} \cos( 2 \omega(\tau_2 + \tau_4+\tau_3) + \theta_2 + \theta_4+\theta_3) \;.
\end{eqnarray}
Still we notice that in this case it is impossible to obtain the identity (\ref{DDF|}) due to the mismatching between the $1/2$ prefactor of $g_1(\omega)$ and the $1/4$ prefactor of $g_2(\omega)$.

\subsection{Exclusive zero-coincidence point for $k>4$} 
From the previous subsections it is clear that as $k$ increases determining  the condition under which Eq.~(\ref{RBP0}) can be fulfilled becomes harder and harder. In particular the absence of a solution for $k=3$ suggests that the problem cannot be solved by induction. Still we suspect that for large enough $k$, the possible combinations of $\theta_2$, $\cdots$, $\theta_k$ ensuring the condition 
\begin{eqnarray}\label{TOTAL0k}
\mbox{Perm}_k(\omega,\omega^\prime)\Big|_{\tau_1=\cdots=\tau_k=0}=0\;,
\end{eqnarray} 
will increase. For instance in passing from $k=3$ to $k=4$ we go from a very constrained set solutions where one among $\theta_2$ and $\theta_3$ is forced to be an integer multiple of $\pi$ (see Eq.~(\ref{COND333})) to (\ref{QUESTA}) which allows $\theta_2$, $\theta_3$ and $\theta_4$ to span over a dense set of possibilities. Exploiting this increased freedom in the selection of  $\theta_2$, $\cdots$, $\theta_k$ it is reasonable to assume that among them it would be possible to find a special assignment to force Eq.~(\ref{RBP0}). Accordingly it is expected that this problem is always solvable for $k\geq 4$ (possibly excluding some isolated and pathological cases).

\begin{figure*}[!t]
	\centering
	\resizebox{0.7\textwidth}{!}{\includegraphics{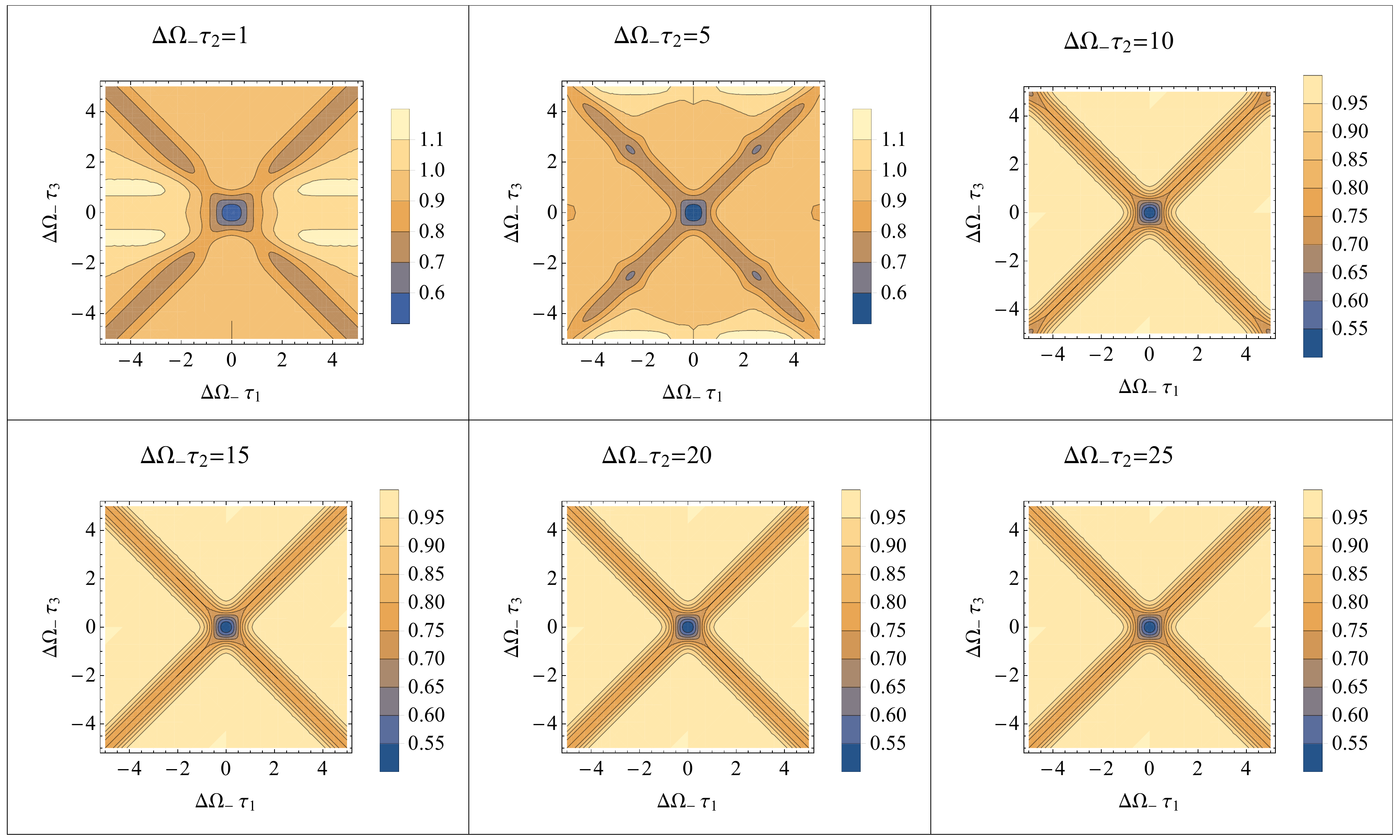}}
	\caption
	{(Color online) Contour plots of the coarse-grained coincidence counts $\bar{R}_{\mbox{\tiny{BP}}}(\tau_1,\tau_2,\tau_3)$ of Eq.~(\ref{RESULT1}) for assigned values of $\tau_2$ in the generalized $k=3$ HOM configuration. All of the time-delays are rescaled by the inverse of the width $\Delta \Omega_-$ of the biphoton frequency-spectrum function (see Eq.~(\ref{GAUS})), and the coincidence counts is rescaled by the plateau value, i.e.  $\bar{R}_{BP}(0, \tau_2, \infty)=1/2$.}
	\label{FIGDELAY13} 
\end{figure*}

\begin{figure*}[!t]
	\centering
	\resizebox{0.7\textwidth}{!}{\includegraphics{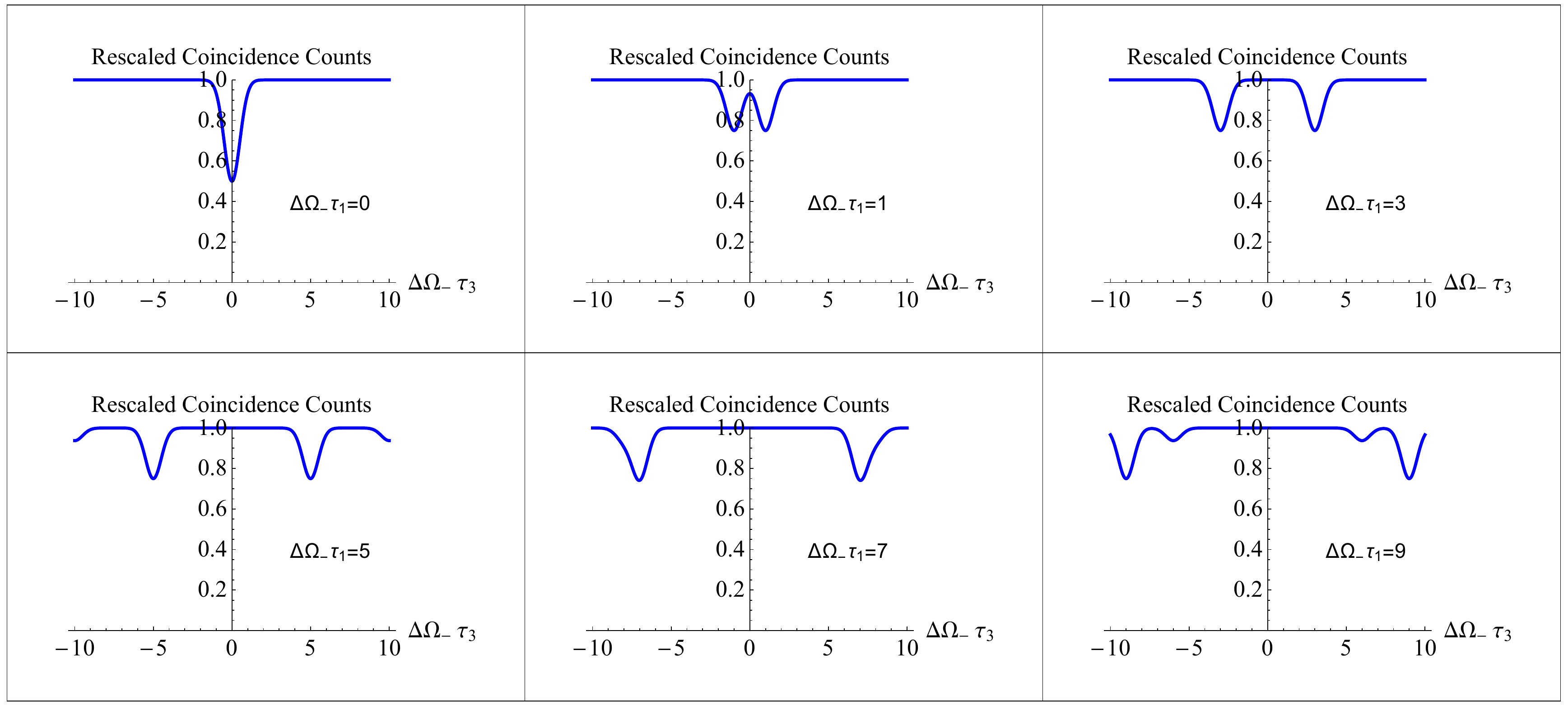}}
	\caption
	{(Color online)  Functional dependence of $\bar{R}_{BP}(\tau_1,15/\Delta \Omega_-,\tau_3)$ of Eq.~(\ref{RESULT1}) upon $\tau_3$ for the assigned values of $\tau_1$. Here the time-delay $\tau_j (j=1,2,3)$ are rescaled by the inverse of the width $\Delta \Omega_-$ of the biphoton frequency-spectrum function. All of the coincidence counts are rescaled by the plateau value, i.e.  $\bar{R}_{BP}(0,15/ \Delta \Omega_-,\infty)=1/2$.}
	\label{DEPENDENCE13} 
\end{figure*}

\begin{figure*}[!t]
	\centering
	\resizebox{0.7\textwidth}{!}{\includegraphics{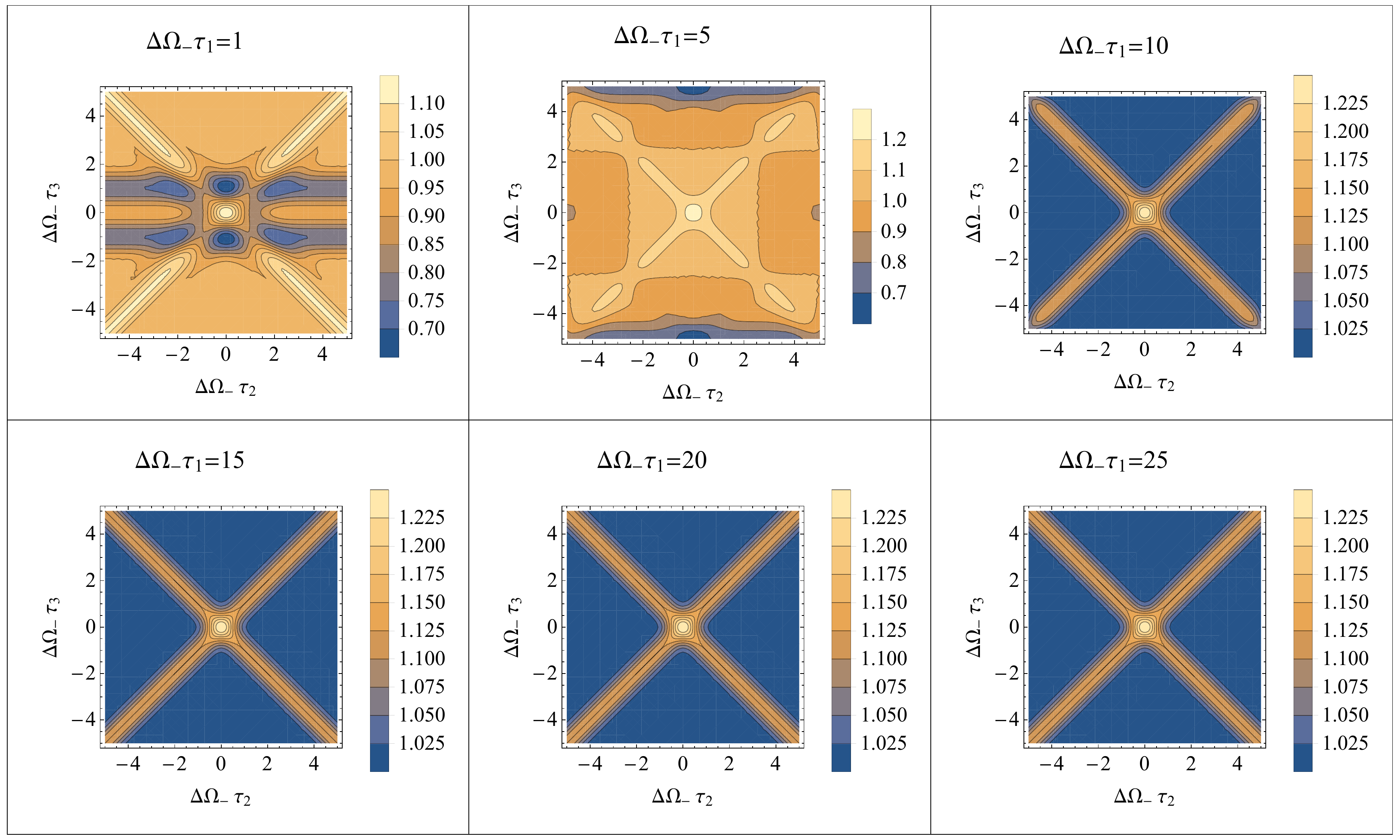}}
	\caption
	{(Color online) Contour plots of the coarse-grained coincidence counts $\bar{R}_{\mbox{\tiny{BP}}}(\tau_1,\tau_2,\tau_3)$ of Eq.~(\ref{RESULT1}) for assigned values of $\tau_1$  in the generalized $k=3$ HOM configuration. All of the time-delays are rescaled by the inverse of the width $\Delta \Omega_-$ of the biphoton frequency-spectrum function (see Eq.~(\ref{GAUS})), and the coincidence counts is rescaled by the plateau value, i.e.  $\bar{R}_{BP}(\tau_1,0,\infty)=1/2$.}
	\label{FIGDELAY23} 
\end{figure*}

\begin{figure*}[!t]
	\centering
	\resizebox{0.7\textwidth}{!}{\includegraphics{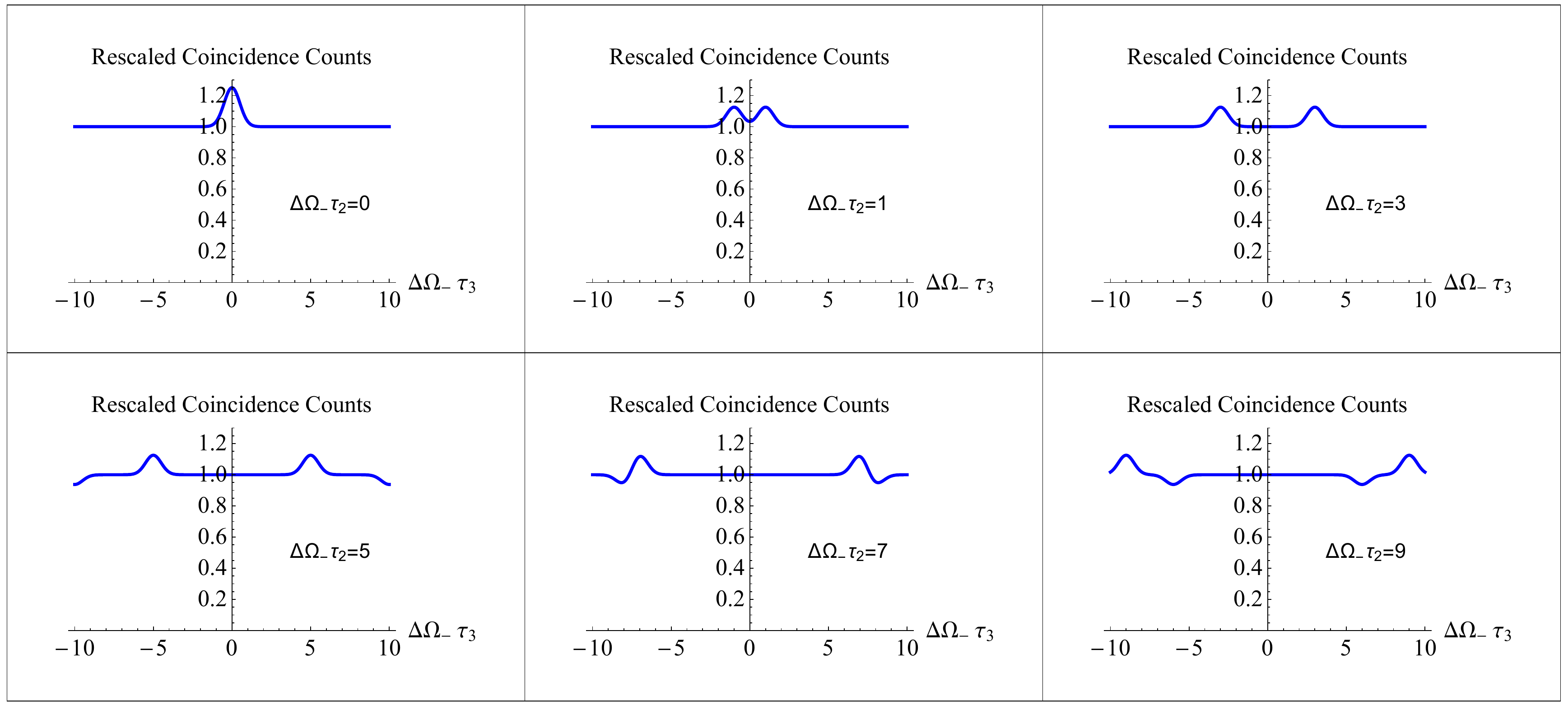}}
	\caption
	{(Color online)   Functional dependence of $\bar{R}_{BP}(15/\Delta \Omega_-,\tau_2,\tau_3)$ of Eq.~(\ref{RESULT1}) upon $\tau_3$ for the assigned values of $\tau_2$. Here the time-delay $\tau_j(j=1,2,3)$ is rescaled by the inverse of the width $\Delta \Omega_-$ of the biphoton frequency-spectrum function. All of the coincidence counts are rescaled by the plateau value $\bar{R}_{BP}(15/ \Delta \Omega_-,0,\infty)=1/2$.}
	\label{DEPENDENCE23} 
\end{figure*}

\section{Sensitivity to fluctuations}\label{sec:sense}
To get a concrete example we now specialize the analysis under the assumption of a BP input state~(\ref{IN}) with Gaussian two-mode spectral function of the form 
\begin{eqnarray}\label{GAUS} 
|\Psi_S(\omega,\omega')|^2 = P_+(\omega+\omega') P_-(\omega-\omega')\;, 
\end{eqnarray} 
with the following normal distributions 
\begin{eqnarray} 
P_+(\omega+\omega')&=&\frac{1}{\sqrt{2\pi} \Delta \Omega_+} e^{ - \frac{(\omega + \omega' - 2 \omega_0)^2}{8 \Delta^2 \Omega_+}}\;,  \nonumber\\
P_-(\omega-\omega')&=&\frac{1}{\sqrt{2\pi} \Delta \Omega_-}  e^{ - \frac{(\omega-\omega')^2}{2 \Delta^2 \Omega_-}}\;, 
\end{eqnarray} 
which locally assigns to each  photon an average  frequency $\omega_0$ with spread $\Delta \omega: =\sqrt{\Delta \Omega_-^2 + 4 \Delta \Omega_+^2}/2$. In the limit $\Delta \Omega_+ \ll \Delta \Omega_-$, Eq.~(\ref{GAUS}) approaches the frequency-entangled biphoton state emerging from an ideal Spontaneous Parametric Down Conversion (SPDC) source pumped with a laser of mean frequency $\omega_p = 2 \omega_0$, e.g.~\cite{BIPHOTON0,BIPHOTON1,BIPHOTON2,BIPHOTON3} and references therein; 
for $\Delta \Omega_+ =\Delta \Omega_-$ instead it corresponds to two uncorrelated (unentangled) single-photon packets; 
while finally for $\Delta \Omega_+ \gtrapprox \Delta \Omega_-$ it mimics the properties of an entangled state emitted by a difference-beam (DB) source~\cite{BP-CP-STATE}.
Under the condition~(\ref{GAUS}) it follows that the associated coincidence counts~(\ref{BP}) can be expressed as the sum of two contributions, 
\begin{eqnarray}\label{BPRE}
&&R^{(\theta_2, \cdots ,\theta_k)}_{\mbox{\tiny{BP}}}(\tau_1,\cdots,\tau_k) \nonumber\\
&&=\int d\omega \int  d\omega^\prime  |  \Psi_S(\omega,\omega^\prime) |^2 |\mbox{Perm}_k(\omega,\omega')|^2  \nonumber\\
&&= \bar{R}_{\mbox{\tiny{BP}}}(\tau_1, \cdots ,\tau_k) + \Delta R^{(\theta_2, \cdots, \theta_k)}_{\mbox{\tiny{BP}}}(\tau_1,\cdots,\tau_k)\;, 
\end{eqnarray} 
where the whole dependence upon $\theta_2$, $\cdots$, $\theta_k$ is only carried by $\Delta R^{(\theta_2, \cdots, \theta_k)}_{\mbox{\tiny{BP}}}(\tau_1,\cdots,\tau_k)$ which, as a function of $\tau_1$, $\cdots$, $\tau_k$, exhibits a series of fast oscillations with frequency $\omega_0$, and where $\bar{R}_{\mbox{\tiny{BP}}}(\tau_1, \cdots ,\tau_k)$ depends instead only upon the spectral width  $\Delta \Omega_-$ which determines the spread of the frequency mismatch of the two input photons (see Sec.~\ref{appc} for details).
The special decoupling enlightened in Eq.(\ref{BPRE}) implies that in the presence of fluctuations of the delays $\tau_1$, $\cdots$, $\tau_k$, the functional dependence of $R^{(\theta_2, \cdots ,\theta_k)}_{\mbox{\tiny{BP}}}(\tau_1,\cdots,\tau_k)$ upon $\theta_2, \cdots, \theta_k$ is washed away. To see this assume that each of these parameters fluctuates randomly and independently on intervals $T$ which are much larger than the inverse of the mean biphoton frequency $\omega_0$ but smaller than the inverse of the associated local spread $\Delta \omega$ (i.e. $1/\omega_0 \ll T \ll 1/\Delta \omega$). In this case the result of coincidence counts registered at the output of the set-up is effectively provided by the coarse graining counterpart of $R^{(\theta_2, \cdots ,\theta_k)}_{\mbox{\tiny{BP}}}(\tau_1,\cdots,\tau_k)$~\cite{MHOM}, i.e. 
\begin{widetext}
\begin{eqnarray}  \label{CORS}
\Big\langle {R}^{(\theta_2, \cdots, \theta_k)}_{\mbox{\tiny{BP}}}(\tau_1,\cdots,\tau_k) \Big\rangle:= \int_{\tau_1-\frac{T}{2}}^{\tau_1+\frac{T}{2}}\frac{ d\tau_1^\prime}{T} \cdots \int_{\tau_k-\frac{T}{2}}^{\tau_k+\frac{T}{2}} \frac{d\tau_k^\prime}{T} \;  R^{(\theta_2, \cdots, \theta_k)}_{\mbox{\tiny{BP}}}(\tau_1^\prime,\cdots, ,\tau_k^\prime)\;.
\end{eqnarray} 
Now from Eq.~(\ref{BPRE}) it  follows that 
\begin{eqnarray}  \label{CORS1}
\Big\langle {R}^{(\theta_2, \cdots, \theta_k)}_{\mbox{\tiny{BP}}}(\tau_1,\cdots,\tau_k) \Big\rangle &\simeq& \bar{R}_{\mbox{\tiny{BP}}}(\tau_1,\cdots,\tau_k)\;, \quad \quad
\Big\langle \Delta R^{(\theta_2,\cdots, \theta_k)}_{\mbox{\tiny{BP}}}(\tau_1^\prime,\cdots,\tau_k^\prime)\Big\rangle  \simeq  0\;,
\end{eqnarray}
\end{widetext}  
(see again Sec.~\ref{appc} for a formal derivation of these identities that holds for arbitrary values of $k$). Eq.(\ref{CORS1}) makes it explicit the fact that the achromatic shifts $\theta_2$, $\cdots$, $\theta_k$, while enforcing the exclusive zero-coincidence point condition~(\ref{RBP0}), become inconsequential in the presence of fluctuations of the time-delays. 
For $k=1$ this is a direct consequence of the fact that in the original HOM scheme one has  
\begin{eqnarray} \label{R11} 
\bar{R}_{\mbox{\tiny{BP}}}(\tau_1)={R}_{\mbox{\tiny{BP}}}(\tau_1)=\frac{1}{2} \Big(1 - e^{- 2  \tau_1^2\Delta \Omega_-^2}\Big)\;, 
\end{eqnarray} 
the contribution $\Delta{R}_{\mbox{\tiny{BP}}}(\tau_1)$ being explicitly zero after the coarse graining -- of course this is also the only case which exhibits a zero-coincidence point without achromatic phase shifters. For the case $k=2$ instead we get~\cite{MHOM}
\begin{eqnarray}  \label{RR2} 
&&\bar{R}_{\mbox{\tiny{BP}}}(\tau_1,\tau_2)=\frac{1}{2}+ \frac{1}{8}\Big( 2 e^{- 2  \tau_2^2\Delta \Omega_-^2} \nonumber\\
&&-e^{- 2 (\tau_1+\tau_2)^2\Delta \Omega_-^2} -  e^{- 2  (\tau_1-\tau_2)^2\Delta \Omega_-^2}\Big)\;,
\end{eqnarray} 
while, as explicitly discussed in Appendix~\ref{PREFACTOR}, for $k=3$ we get
\begin{widetext}
\begin{eqnarray}\label{RESULT1}
	\bar{R}_{\mbox{\tiny{BP}}}(\tau_1,\tau_2,\tau_3) &=& \frac{1}{2}
	+\frac{1}{32} \Big(2 e^{-2(\tau_2-\tau_3)^2\Delta\Omega_-^2}+2 e^{-2(\tau_2+\tau_3)^2\Delta\Omega_-^2}-
	4 e^{-2(\tau_1-\tau_3)^2\Delta\Omega_-^2}- 4 e^{-2(\tau_1+\tau_3)^2\Delta\Omega_-^2} \nonumber \\ && 
	-e^{-2(\tau_1+\tau_2-\tau_3)^2\Delta\Omega_-^2}-e^{-2(\tau_1-\tau_2+\tau_3)^2\Delta\Omega_-^2}-e^{-2(\tau_1-\tau_2-\tau_3)^2\Delta\Omega_-^2}-e^{-2(\tau_1+\tau_2+\tau_3)^2\Delta \Omega_-^2}\Big) \;,
\end{eqnarray} 
which we plot in Figs.\ref{FIGDELAY13}--\ref{DEPENDENCE23}.
\end{widetext}

\subsection{Effect of the coarse graining} \label{appc} 
Here we present a derivation of Eq.~(\ref{CORS1}) for arbitrary $k$. We start observing  that from Eq.~(\ref{DEFNk}) it follows that the various matrix elements of $N_k(\omega)$ can be expressed as 
\begin{eqnarray} 
A_k(\omega):= \sum_{\vec{s}}  \alpha_k^{(\vec{s})}  e^{ i  \vec{s} \cdot \vec{\varphi}(\omega)}\;, \quad
B_k(\omega):= \sum_{\vec{s}}  \beta_k^{(\vec{s})}  e^{ i  \vec{s} \cdot \vec{\varphi}(\omega)} \;, \nonumber\\
C_k(\omega):= \sum_{\vec{s}}  \gamma_k^{(\vec{s})}  e^{ i  \vec{s} \cdot \vec{\varphi}(\omega)}\;, \quad
D_k(\omega):= \sum_{\vec{s}}  \delta_k^{(\vec{s})}  e^{ i  \vec{s} \cdot \vec{\varphi}(\omega)}\;,  \nonumber\\
\end{eqnarray} 
where the summation runs over the set of $k$-long vector ${\vec{s}}$ formed by the sequences of $\pm 1$, where $\vec{\varphi}(\omega):= (\varphi_1(\omega), \cdots, \varphi_k(\omega))$, and where finally  $\alpha_k^{(\vec{s})}$, $\beta_k^{(\vec{s})}$, $\gamma_k^{(\vec{s})}$, and $\delta_k^{(\vec{s})}$ are the complex parameters of modulus $2^{k/2}$ which are independent from the phase $\varphi_\ell(\omega)$.  Accordingly we can write 
\begin{eqnarray} \label{impor1}
&&\mbox{Perm}_k(\omega,\omega') \nonumber\\
&=& \sum_{\vec{s},\vec{s'}} [ \alpha_k^{(\vec{s})} \delta_k^{(\vec{s}')} + \beta_k^{(\vec{s}')} \gamma_k^{(\vec{s})} ] \; e^{ i  [\vec{s} \cdot \vec{\varphi}(\omega) + \vec{s}'\cdot \vec{\varphi}(\omega')] } \nonumber \\   
&=& \sum_{\vec{s},\vec{s'}} Q_k^{(\vec{s},\vec{s'})} \; e^{ i  [(\omega+\omega') \; \vec{\tau} +\vec{\theta} ] \cdot \vec{\Delta}_+ (\vec{s},\vec{s}') } \; e^{ i (\omega-\omega') \; \vec{\tau}\cdot \vec{\Delta}_- (\vec{s},\vec{s}') } \;,  \nonumber\\
\end{eqnarray} 
where in the second identity we  introduced the quantities
\begin{eqnarray}
&&Q_k^{(\vec{s},\vec{s'})}:= [\alpha_k^{(\vec{s})} \delta_k^{(\vec{s}')} + \beta_k^{(\vec{s}')} \gamma_k^{(\vec{s})}] \;, \\
&&\vec{\Delta}_\pm(\vec{s},\vec{s}'):=({ \vec{s}\pm \vec{s'} })/{2}\;, 
\end{eqnarray}
and used Eq.~(\ref{PHASESHIFT})  to write 
\begin{eqnarray}
&&\vec{s} \cdot \vec{\varphi}(\omega) + \vec{s}'\cdot \vec{\varphi}(\omega')    \nonumber\\
&&=[(\omega+\omega') \; \vec{\tau} +\vec{\theta} ] \cdot \vec{\Delta}_+ (\vec{s},\vec{s}')+(\omega-\omega') \; \vec{\tau}\cdot \vec{\Delta}_- (\vec{s},\vec{s}') \;,   \nonumber\\
\end{eqnarray} 
with $\vec{\tau}:=(\tau_1,\cdots, \tau_k)$ and $\vec{\theta}:=(\theta_1,\cdots, \theta_k)$. 
We now split the summation of Eq.~(\ref{impor1}) into $\vec{s}=-\vec{s}'$ term and $\vec{s} \ne -\vec{s}'$ term respectively, i.e.
\begin{eqnarray} \label{FID} 
&&\mbox{Perm}_k(\omega,\omega')=F_k(\omega-\omega')  \nonumber\\
&&+\sum_{\vec{s}\neq -\vec{s'}} Q_k^{(\vec{s},\vec{s'})} \; e^{ i  [(\omega+\omega') \; \vec{\tau} +\vec{\theta} ] \cdot \vec{\Delta}_+ (\vec{s},\vec{s}') } \;  e^{ i (\omega-\omega') \; \vec{\tau}\cdot \vec{\Delta}_- (\vec{s},\vec{s}') }\;, \nonumber\\
\end{eqnarray} 
with
\begin{eqnarray}
F_k(\omega-\omega') : =\sum_{\vec{s}} Q_k^{(\vec{s},-\vec{s})} \; e^{ i (\omega-\omega') \; \vec{\tau}\cdot \vec{s} }\;, 
\end{eqnarray}
where we use the fact that $\vec{\Delta}_- (\vec{s},-\vec{s})=\vec{s}$ and $\vec{\Delta}_+ (\vec{s},-\vec{s})=0$. Taking the modulus we have
\begin{widetext}
\begin{eqnarray} \label{FID1} 
|\mbox{Perm}_k(\omega,\omega')|^2 
&=& |F_k(\omega-\omega')|^2 +  \Big| \sum_{\vec{s}\neq -\vec{s'}} Q_k^{(\vec{s},\vec{s'})}
e^{ i  [(\omega+\omega') \; \vec{\tau} +\vec{\theta} ] \cdot \vec{\Delta}_+ (\vec{s},\vec{s}') } \; 
e^{ i (\omega-\omega') \; \vec{\tau}\cdot \vec{\Delta}_- (\vec{s},\vec{s}') } \Big|^2 \nonumber \\
&&+ 2  \mbox{Re}\Big[  F^*_k(\omega-\omega')  \sum_{\vec{s}\neq -\vec{s'}} Q_k^{(\vec{s},\vec{s'})}
e^{ i  [(\omega+\omega') \; \vec{\tau} +\vec{\theta} ] \cdot \vec{\Delta}_+ (\vec{s},\vec{s}') } \; 
e^{ i (\omega-\omega') \; \vec{\tau}\cdot \vec{\Delta}_- (\vec{s},\vec{s}') } \Big] \;, 
\end{eqnarray}
\end{widetext}
which is further expressed as 
\begin{eqnarray} \label{FID11} 
|\mbox{Perm}_k(\omega,\omega')|^2  = \bar{P}_k(\omega-\omega') + \Delta P_k(\omega,\omega')\;, 
\end{eqnarray} 
with $\bar{P}_k(\omega-\omega')$ being a function that only depends upon $\omega-\omega'$ but bares no functional dependence neither upon $\omega+\omega'$ nor upon $\vec{\theta}$, while $\Delta P_k(\omega,\omega')$ being a sum of contributions which exhibit the phase-shift terms that have a non-trivial dependence upon $\omega+\omega'$. Specifically,  
\begin{widetext}
\begin{eqnarray} \label{PPPD} 
\bar{P}_k(\omega-\omega') &:=& |F_k(\omega-\omega')|^2 +   \sum_{\vec{s}\neq -\vec{s'}} |Q_k^{(\vec{s},\vec{s'})}|^2  + 2  \mbox{Re}\Big[  \sum_{\{ \vec{s}\neq \vec{s'},\vec{s}_1\neq \vec{s'}_1\} \in {\cal S}_+}  Q_k^{(\vec{s},\vec{s'})} Q_k^{(\vec{s}_1,\vec{s'}_1)*} \;  e^{ i (\omega-\omega') \; \vec{\tau}\cdot [\vec{\Delta}_- (\vec{s},\vec{s}') -  \vec{\Delta}_- (\vec{s}_1,\vec{s}'_1)] } \Big] \;, \nonumber \\ 
\Delta P_k(\omega,\omega')&:=& 2\mbox{Re}\Big[   \sum_{\{ \vec{s}\neq \vec{s'},\vec{s}_1\neq \vec{s'}_1\} \in {\cal S}_-}   Q_k^{(\vec{s},\vec{s'})} Q_k^{(\vec{s}_1,\vec{s'}_1)*} e^{ i [(\omega+\omega') \; \vec{\tau}+\vec{\theta}]\cdot [\vec{\Delta}_+ (\vec{s},\vec{s}') -  \vec{\Delta}_+ (\vec{s}_1,\vec{s}'_1)] } \Big]  \; \nonumber \\
&&+ 2\mbox{Re}\Big[  F^*_k(\omega-\omega')  \sum_{\vec{s}\neq -\vec{s'}} Q_k^{(\vec{s},\vec{s'})} e^{ i  [(\omega+\omega') \; \vec{\tau} +\vec{\theta} ] \cdot \vec{\Delta}_+ (\vec{s},\vec{s}') } \; e^{ i (\omega-\omega') \; \vec{\tau}\cdot \vec{\Delta}_- (\vec{s},\vec{s}') } \Big]\;, 
\end{eqnarray} 
\end{widetext}
with  ${\cal S}_{+}$ and ${\cal S}_{-}$  being the subsets formed by the couples $\{ \vec{s}\neq \vec{s'}\} \neq\{\vec{s}_1\neq \vec{s'}_1\}$ such that $\vec{\Delta}_+ (\vec{s},\vec{s}') = \vec{\Delta}_+ (\vec{s}_1,\vec{s}'_1)$ and $\vec{\Delta}_- (\vec{s},\vec{s}') = \vec{\Delta}_- (\vec{s}_1,\vec{s}'_1)$ respectively (notice that they have zero overlap). 
Using Eq.~(\ref{FID11}) to compute the coincidence counts we hence arrive to Eq.~(\ref{BPRE}) via the identification 
\begin{eqnarray} \label{DDs1} 
&&\bar{R}_{\mbox{\tiny{BP}}}(\tau_1, \cdots ,\tau_k) \nonumber\\
&&=\int d\omega \int  d\omega^\prime  |  \Psi_S(\omega,\omega^\prime) |^2 \bar{P}_k(\omega-\omega') \nonumber\\
&&=\int d\omega \int  d\omega^\prime  \; P_+(\omega+\omega') P_-(\omega-\omega') \bar{P}_k(\omega-\omega')\;,
\end{eqnarray} 
and 
\begin{eqnarray} \label{DDs2} 
&&\Delta R^{(\theta_2, \cdots, \theta_k)}_{\mbox{\tiny{BP}}}(\tau_1,\cdots,\tau_k)  \nonumber\\
&&=\int d\omega \int  d\omega^\prime  |  \Psi_S(\omega,\omega^\prime) |^2 \Delta{P}_k(\omega,\omega')  \nonumber\\
&&=\int d\omega \int  d\omega^\prime  P_+(\omega+\omega') P_-(\omega-\omega')\Delta{P}_k(\omega,\omega')\;. 
\end{eqnarray}
We used (\ref{GAUS}) to split the integral in terms of the variable $\xi:=\omega+\omega'$ and $\nu:=\omega-\omega'$, hence Eq.~(\ref{DDs2}) is given by
\begin{eqnarray}
&&\left( \frac{1}{2} \int d\xi   P_+(\xi) e^{ i  [\xi \vec{\tau} +\vec{\theta} ] \cdot \vec{K}_+} \right) \left( \int  d\nu P_-(\nu) e^{ i\nu \; \vec{\tau}\cdot \vec{K}_- }\right) \nonumber\\
&&=e^{ i  (2\omega_0\vec{\tau} +\vec{\theta})  \cdot \vec{K}_+} e^{- 2 (\vec{\tau}\cdot \vec{K}_+)^2 \Delta \Omega_+^2} e^{-{ (\vec{\tau}\cdot \vec{K}_-)^2 \Delta \Omega_-^2}/{2}} \;,
\end{eqnarray} 
with the vector $\vec{K}_+$ being explicitly non-zero. As anticipated in the main text before, $\Delta R^{(\theta_2, \cdots, \theta_k)}_{\mbox{\tiny{BP}}}(\tau_1,\cdots,\tau_k)$ depends explicitly upon $\omega_0$ and $\vec{\theta}$. Furthermore we notice that under coarse graining we have 
\begin{eqnarray} \label{AVE1}
&&\Big\langle e^{ i  (2\omega_0\vec{\tau} +\vec{\theta})  \cdot \vec{K}_+}e^{- 2 (\vec{\tau}\cdot \vec{K}_+)^2 \Delta \Omega_+^2} e^{-{ (\vec{\tau}\cdot \vec{K}_-)^2 \Delta \Omega_-^2}/{2}}\Big\rangle  \nonumber\\
&&\simeq \Big\langle e^{ i  (2\omega_0\vec{\tau} +\vec{\theta})  \cdot \vec{K}_+} \Big\rangle  \simeq  0\;, 
\end{eqnarray} 
where in the first identity we used the fact that since we assume $1/\omega_0 \ll T \ll 1/\Delta \omega$, the integral is performed over intervals of length $T$ which is much shorter than $1/\Delta \Omega_-$ and $1/\Delta \Omega_+$. Inserting this into (\ref{DDs2}) we can thus conclude that 
\begin{eqnarray}
\Big\langle \Delta R^{(\theta_2,\cdots, \theta_k)}_{\mbox{\tiny{BP}}}(\tau_1^\prime,\cdots,\tau_k^\prime)\Big\rangle  \simeq  0\;,
\end{eqnarray} 
which proves the second identity of Eq.~(\ref{CORS1}). The first one follows along the same line by observing that according to Eq.~(\ref{PPPD}) the right-hand-side term of (\ref{DDs1}) is given by a finite summation of terms which are either constant or have the following dependence on $\vec{\tau}$, 
\begin{eqnarray}\label{DDS1} 
&&\left( \frac{1}{2} \int d\xi   P_+(\xi) \right) \left(\int  d\nu P_-(\nu)\; e^{ i\nu \; \vec{\tau}\cdot \vec{K}_- }\right) \nonumber\\
&&= e^{-{ (\vec{\tau}\cdot \vec{K}_-)^2 \Delta \Omega_-^2}/{2}} \;,
\end{eqnarray} 
with $\vec{K}_-$ being some non-zero vectors, we notice that the values of $\bar{R}_{\mbox{\tiny{BP}}}(\tau_1)$, $\bar{R}_{\mbox{\tiny{BP}}}(\tau_1,\tau_2)$, and $\bar{R}_{\mbox{\tiny{BP}}}(\tau_1,\tau_2,\tau_3)$ reported in Eqs.~(\ref{R11})--(\ref{RESULT1}) have indeed this structure. Therefore, repeating the same argument used in Eq.(\ref{AVE1}) in this case we get
\begin{eqnarray}
\Big\langle e^{-{ (\vec{\tau}\cdot \vec{K}_-)^2 \Delta \Omega_-^2}/{2}} \Big\rangle\simeq e^{-{ (\vec{\tau}\cdot \vec{K}_-)^2 \Delta \Omega_-^2}/{2}}\;, 
\end{eqnarray}  
which ultimately leads to 
\begin{eqnarray}  \label{CORS1123}
\Big\langle \bar{R}_{\mbox{\tiny{BP}}}(\tau_1,\cdots,\tau_k) \Big\rangle  \simeq  \bar{R}_{\mbox{\tiny{BP}}}(\tau_1,\cdots,\tau_k)\;,
\end{eqnarray} 
and hence get the first identity of Eq.~(\ref{CORS1}).

\section{Conclusion}\label{CONCS}
A generalized multi-parameter HOM interferometer composed by $k$ 50:50 beam splitters, $k$ different time-delays and $(k-1)$ achromatic wave-plates has been presented. In the special case with $k=2$ modules, the described setup was employed in Ref.~\cite{MHOM} as a scheme to measure two independent time-delays parameters via the results of coincidence counts at the output. Borrowing directly from the original HOM scheme~\cite{HOM1}, the idea was to link the uniqueness of a HOM zero-coincidence point attainable when setting $\theta_2=\pi/2$, to a way for detecting the non-zero values of $\tau_1$ and $\tau_2$, thereby compensating for them by adding the controllable delays along the interferometric paths. From the results presented at here, it is clear that the same construction can be extended to the case of $k=4$ modules, while this would not be possible for the case of just three modules due to the impossibility of fulfilling the condition (\ref{RBP0}) for $k=3$. In Ref.\cite{MHOM} it was also shown how to use the residual functional dependence of the coarse-grained coincidence counts~(\ref{RR2}) upon the delays to determine their values. Clearly the same construction can be applied also for larger values of $k$. In particular this can be done for the unlucky case $k=3$ which, in the presence of fluctuations, does not even admits an exclusive HOM zero-coincidence point. As a matter of fact, from Figs.\ref{FIGDELAY13}--\ref{DEPENDENCE13} we notice that fixing $\Delta \Omega_-\tau_2 \ge 10$, $\bar{R}_{\mbox{\tiny{BP}}}(\tau_1,\tau_2,\tau_3)$ exhibits two symmetric dips as a function of $\tau_3$ for assigned values of $\tau_1$, with a visibility $V_{min} \approx 25\%$ (the ratio between the depth of the minima and the height of the plateau). On the contrary, from Figs.\ref{FIGDELAY23}--\ref{DEPENDENCE23} it follows that for $\Delta \Omega_-\tau_1 \ge 10$, $\bar{R}_{\mbox{\tiny{BP}}}(\tau_1,\tau_2,\tau_3)$ exhibits instead two symmetric peaks as a function of $\tau_3$ for assigned values of $\tau_2$, with the visibility $V_{max} \approx 10\%$ (the ratio between  the maximum value and the height of the plateau). Consider then the case where each of the length-difference of paths $\mathbf{A}_1\mathbf{B}_1$, $\mathbf{A}_2\mathbf{B}_2$ and $\mathbf{A}_3\mathbf{B}_3$ as $\Delta \ell_j=\Delta\ell_j^{(0)}+x_j$, $\forall j=1,2,3$, where $\Delta\ell_j^{(0)}$ is fixed and unknown, the second term $x_j$ is controllable by the experimentalist.
One way to recover these parameters could be the following procedure: i) we select $x_2$ to be sufficiently large to ensure that the value  $\Delta \tau_2=\Delta \ell_2/2c$ to be larger than $10/\Delta \Omega_-$. Then keeping $x_1=0$, we record the values of $\bar{R}_{\mbox{\tiny{BP}}}(\frac{\Delta\ell_1^{(0)}}{2c}, \frac{\Delta \ell_2}{2c},\frac{\Delta\ell_3^{(0)}+x_3}{2c})$ as a function of $x_3$, locating the two minima $x_{min}^{(r)}$, $x_{min}^{(l)}$ of Fig.\ref{DEPENDENCE13}. This allows us to determine the values of $\Delta \ell_1^{(0)}$ and $\Delta \ell_3^{(0)}$ by observing that $\Delta \ell_1^{(0)}=x_{min}^{(r)}-x_{min}^{(l)}$ and $\Delta \ell_3^{(0)}=x_{min}^{(r)}+x_{min}^{(l)}$ respectively; 
ii) With this information we now set $x_1$ to get $\Delta \tau_1=\Delta \ell_1/2c$ larger than $10/\Delta \Omega_-$, keeping $x_2=0$ and start scanning once more $\bar{R}_{\mbox{\tiny{BP}}}(\frac{\Delta\ell_1}{2c}, \frac{\Delta \ell^{(0)}_2}{2c},\frac{\Delta\ell_3^{(0)}+x_3}{2c})$ with respect to $x_3$ to locate the maxima $x_{max}^{(r)}$, $x_{max}^{(l)}$ of Fig.\ref{DEPENDENCE23}, hence the value of $\Delta \ell_2^{(0)}$ can be obtained as $\Delta \ell_2^{(0)}=x_{max}^{(r)}-x_{max}^{(l)}$.

The Authors acknowledge support from the Grant \textit{National Natural Science Foundation of China 61172138, 61573059, 61771371; Aviation Science Foundation Project 20160181004; Shanghai Aerospace Science and Technology Innovation Fund SAST2017-030}, and SNS for hosting her.

\appendix 
\section{Appendix} \label{PREFACTOR}
As for the $k=3$ modules, $\Delta R^{(\theta_2, \theta_3)}_{\mbox{\tiny{BP}}}(\tau_1,\tau_2,\tau_3)$ of Eq.(\ref{BPRE}) with respect to $\theta_2=\pi/2$ and $\theta_3=0$ is given by
\begin{widetext}
\begin{eqnarray} 
	\Delta R^{(\theta_2, \theta_3)}_{\mbox{\tiny{BP}}}(\tau_1,\tau_2,\tau_3) &=
	&\frac{1}{32}\Big\{f_1(\tau_1,\tau_2,\tau_3)\cos(4\omega_0\tau_2)+f_2(\tau_1,\tau_2,\tau_3)\sin(2\omega_0\tau_2)+f_3(\tau_1,\tau_2,\tau_3)\cos(4\omega_0\tau_3)\nonumber\\
	&&+f_4(\tau_1,\tau_2,\tau_3)[e^{-16\tau_2\tau_3\Delta\Omega_+^2}\cos(4\omega_0(\tau_2+\tau_3))+e^{16\tau_2\tau_3\Delta\Omega_+^2}\cos(4\omega_0(\tau_2-\tau_3))]\nonumber\\
	&&+f_5(\tau_1,\tau_2,\tau_3)[e^{8\tau_2\tau_3\Delta \Omega_+^2}\sin(2\omega_0(\tau_2-2\tau_3))-e^{-8\tau_2\tau_3\Delta \Omega_+^2}\sin(2\omega_0(\tau_2+2\tau_3))]\Big\}\;,
\end{eqnarray}
with
\begin{eqnarray}
f_1(\tau_1,\tau_2,\tau_3)&=&2\text{exp}\Big[ {-8\tau_2^2\Delta \Omega_+^2-2(\tau_1+\tau_3)^2\Delta \Omega_-^2} \Big]  (1+e^{8\tau_1\tau_3\Delta \Omega_-^2}+2e^{(2\tau_1^2+4\tau_1\tau_3)\Delta \Omega_-^2})\;,\nonumber\\
f_2(\tau_1,\tau_2,\tau_3)&=&-4\text{exp}\Big[ {-2\tau_2(\tau_1+\tau_3)\Delta \Omega_-^2-2(\tau_1+\tau_3)^2\Delta \Omega_-^2 -\frac{1}{2}\tau_2^2(4\Delta \Omega_+^2+\Delta \Omega_-^2)} \Big]  \nonumber\\
&&\times(e^{4\tau_3(2\tau_1+\tau_2)\Delta \Omega_-^2}-e^{4\tau_2(\tau_1+\tau_3)\Delta \Omega_-^2}+e^{4\tau_1(\tau_2+2\tau_3)\Delta \Omega_-^2}-1)\;,\nonumber\\
f_3(\tau_1,\tau_2,\tau_3)&=&2\text{exp} \Big[ {-8\tau_3^2\Delta\Omega_+^2-2(\tau_1+\tau_2)^2\Delta\Omega_-^2}\Big] \nonumber\\
&&\times(e^{8\tau_1\tau_2\Delta\Omega_-^2}-4e^{2\tau_2(2\tau_1+\tau_2)\Delta\Omega_-^2}-2e^{2\tau_1(\tau_1+2\tau_2)\Delta\Omega_-^2}+1)\;,\nonumber\\
f_4(\tau_1,\tau_2,\tau_3)&=&-2\text{exp}\Big[ {-2(4\tau_2^2\Delta\Omega_+^2+4\tau_3^2\Omega_+^2+\tau_1^2\Delta\Omega_-^2)}\Big] (1+e^{2\tau_1^2\Delta\Omega_-^2})\;,\nonumber\\
f_5(\tau_1,\tau_2,\tau_3)&=&4 \text{exp}\Big[ {-2(\tau_2^2+4\tau_3^2)\Delta \Omega_+^2-\frac{1}{2}(2\tau_1+\tau_2)^2\Delta \Omega_-^2} \Big] (e^{4\tau_1\tau_2\Delta\Omega_-^2}-1)\;.
\end{eqnarray}
\end{widetext}

\end{document}